\documentclass[aps,pre,preprint,showpacs,preprintnumbers,amsmath,amssymb,11pt]{revtex4}
\usepackage{graphicx}
\usepackage{epsfig}
\usepackage{dcolumn}
\usepackage{bm}

\begin{document}

\title{Entropy production of entirely diffusional Laplacian transfer and the possible role of fragmentation of the boundaries}
\author{ ${\text{K}}{\text{. Karamano}}{{\text{s}}^{1,2}}$}
\author{${\text{S}}{\text{.I}}{\text{. Mistakidi}}{{\text{s}}^1}$}
\author{${\text{T.J}}{\text{. Massar}}{{\text{t}}^3}$}
\author{${\text{I.S}}{\text{. Mistakidi}}{{\text{s}}^4}$}
\affiliation{${}^1{\text{Department of Physics}}{\text{, University
of Athens}}{\text{, Panepistimiopolis}}
{\text{, GR 15784}}{\text{, Athens}}{\text{, Greece}}$,\\
${}^2{\text{Complex Systems Group, Institute of Nuclear and Particle
Physics}},$\\
${}{{\text{NRCPS Demokritos,GR 15310, Aghia Paraskevi
Attiki, Greece}}}$,\\
${}^3{\text{Building, Architecture and Town Planning Department, 194/2}}{\text{,Universite Libre de Bruxelles, (ULB)}}$,\\
${}^4{\text{Hellenic Army Academy Athens}}{\text{, Greece}}$}

\date{\today}

\begin{abstract}
The entropy production and the variational functional of a Laplacian
diffusional field around the first four fractal iterations of a
linear self-similar tree (von Koch curve) is studied analytically
and detailed predictions are stated. In a next stage, these
predictions are confronted with results from numerical resolution of
the Laplace equation by means of Finite Elements computations. After
a brief review of the existing results, the range of distances near
the geometric irregularity, the so-called "Near Field", a situation
never studied in the past, is treated exhaustively. We notice here
that in the Near Field, the usual notion of the active zone
approximation introduced by Sapoval et al.$^{1,2}$ is strictly
inapplicable. The basic new result is that the validity of the
active-zone approximation based on irreversible thermodynamics is
confirmed in this limit, and this implies a new interpretation of
this notion for Laplacian diffusional fields.\\

Keywords: Diffusion; Entropy Production; Fractals; Near Field; Laplacian Fields.
\end{abstract}

\pacs{65.40.Gr, 05.70.Ln, 5.45.Df, 47.53.+n}

\maketitle

\section{Introduction}

On the past decade, an important field of interdisciplinary research
has been developed based on new geometry concepts, namely "the
fractal" geometry of nature, pioneered by Benoit B. Mandelbrot. The
fractal geometry approximates well enough the naturally discovered
disordered morphologies. For instance, the terminal part of the
respiratory system of mammals, biological membranes, porous
electrodes or catalysts, provide some characteristic examples$^{3 -
12}$.

This implies a direct link to Laplacian transfer, as one could
mention that under ordinary conditions many familiar transport
phenomena such as diffusion and heat conduction are described in the
steady state by Laplacian fields. The Laplace equation is
a linear equation, however the complexity of the problem
is due to the geometrical irregularities of the boundaries. Due to
this fact, an important number of studies on the distribution of
Laplacian fields around geometrically irregular - and eventually
fractal - objects, have been justified$^{1, 13 - 18}$.

An old mathematical conjecture states that the Laplacian field
around a fractal object in two dimensions is itself of multifractal
nature. This conjecture, after Makarov's theorem is a mathematical fact.
In a paper considered classical by now, B.B. Mandelbrot
explored for the first time in detail  the Laplacian field around a
deterministic self-similar tree with his famous "zebroide"
(logarithmic) representation$^{19}$. This work was the
starting point for many authors who examined not only
diffusion but also reaction-diffusion systems$^{2, 20 - 27}$.

Also, specialists from mathematics (harmonic analysis) presented
landmark studies on the distribution of harmonic measure around
irregular boundaries. In particular, N.G. Makarov proved a theorem
stating that, whatever the shape of an irregular (simply connected)
boundary in two dimensions might be, the active zone in which most
of the flux generated by a Laplacian field is concentrated, scales
as a length$^{28 - 29}$. More importantly, this result holds exactly
also for the "mathematical" fractal object. In the "multifractal"
picture of the Laplacian field this amounts to the equality of the
first multifractal exponent (the "information dimension") exactly to
the value 1. Later on,  B. Sapoval proposed another interpretation
of Makarov's theorem on the grounds of "information
ensemble"$^{23,30}$. Moreover, the active zone approximation tells
us simply that some parts  of an irregular surface are not really
contributing, remaining "passive", so that all thermodynamic
properties can be estimated by considering only the "active zone".
The notion of the active zone received ample experimental support
through electrochemistry experiments leading to its visualization
with electrodeposition of copper$^{30}$ (optical absorption).
According to the theorems mentioned above, for a Laplacian field the
active zone is always a good approximation. J. Bourgain generalized
this result in higher dimensions$^{3}$, but with a scaling exponent
which depends on the dimensionality of space and is not necessarily
an integer.

Continuing in the same lines, Makarov and Jones proved that the
harmonic measure around some self-similar objects is multifractal.
Recently, due mainly to work by Sapoval's group, the problem of the
multifractal spectrum around von Koch's curve in two (and then in
three) dimensions has been reconsidered on the grounds of
hierarchical random walkers. Important evidence that in three
dimensions there is a nontrivial behavior of the information measure
and a non-integer dimensionality equal to 2.007 is given by Grebenkov
et al.$^{31}$.

In$^{32}$ an alternative way to characterize the complexity of
Laplacian transport across irregular boundaries was proposed, based
on irreversible thermodynamics. More specifically, the focus was set
on the dissipation generated by the underlying process, as the
irregularity of the boundary is increased. Dissipation is measured
by the entropy production, arguably the central quantity of
irreversible thermodynamics. The correct measure of complexity in
this case is the entropy production per unit surface. In that study,
as well as in$^{33}$, attention was followed to the "Far Field",
that is in distances from the object which far exceed the length of
the minimal irregularity of the surface (the cut-off threshold).

The main purpose of this work is firstly to revisit and generalize
the results of$^{32}$ and secondly to explore the behavior of the
entropy production functional in the region of the "Near Field",
i.e. for distances between the membranes (or electrodes) which are
smaller than the smallest detail of the prefractal iteration. It
should also be noticed that the chart of the spatial distribution of
the local entropy production for an increasingly irregular object in
the Near Field does not exist anywhere in the literature to the best
of our knowledge.

The paper is organized as follows. General considerations and
detailed predictions for the active zone concept of an entirely
diffusional field are presented in Sec.~II. In Sec.~III the near
domain predictions are examined in detail and in Sec.~IV these
predictions are corroborated by numerical results obtained with
Finite Elements and the chart of the local entropy production is
portrayed. Finally, in Sec.~V we draw the main conclusions and we
discuss some more general issues.

\section{The Active Zone of  Diffusion  Fields: General considerations for the "Far" and "Near Domain"}

The simplest setting in which possible effects of complex geometry
on entropy production can be identified is given in Figure 1. A
concentration difference ${c_1} - {c_2}$  is applied across the
vertical boundaries of a cell, over a characteristic length ${L_y}$.
The cell obeys to zero flux boundary conditions along the horizontal
direction $x$, but there is now a geometric irregularity consisting
of extending the horizontal characteristic length ${L_x}$  by a bump
in the middle.

Under the condition that $c$ satisfies in the steady state Laplace
equation
\begin{equation}
\label{eq:1}{\nabla  ^2}c = 0,
\end{equation}
it was shown$^{32}$ that, on the grounds of nonequilibrium
thermodynamics, the full expression for the entropy production reads
\begin{equation}
\label{eq:2}P = Dk\int\limits_V {\frac{{{{(\nabla
c)}^2}}}{c}d\textbf{r} },
\end{equation}
while the variational functional which is extremal in the steady state, is given by
\begin{equation}
\label{eq:3}Q = \int\limits_V {{{(\nabla c)}^2}d\textbf{r} }.
\end{equation}
Before addressing the role of complex boundaries in the entropy
production and the variational functional, we evaluate both $c$ and
$P$ in the simple reference case of a two-dimensional box of height
$L{}_y$ and length $L{}_x$. The corresponding solution reads
\begin{equation}
\label{eq:4}c(y) = \frac{{{c_1} - {c_2}}}{{{L_y}}}y + {c_2},
\end{equation}
where ${c_1} = {c_1}({L_y})$ and ${c_2}$ = $c_{2}(0)$. In addition,
the entropy production can be calculated exactly, yielding
\begin{equation}
\label{eq:5}P = Dk\frac{{{L_x}}}{{{L_y}}}({c_2} - {c_1})\ln
\frac{{{c_2}}}{{{c_1}}},
\end{equation}
whereas the variational functional in this case reads
\begin{equation}
\label{eq:6}Q = \frac{{{L_x}}}{{{L_y}}}{({c_2} - {c_1})^2}.
\end{equation}

For the setting of Figure 1, we have shown that in the "Far field"
the entropy production would keep the structure of eq.(5) as far as
y-dependence goes, but the proportionality factor ${L_x}$ would be
modulated based on Makarov's theorem by an additional
size-independent factor $A$ depending on geometry and on
concentrations only
\begin{equation}
\label{eq:7}P = DkA({c_1},{c_2};g){L_x}\frac{1}{{{L_y}}}({c_2} -
{c_1})\ln \frac{{{c_2}}}{{{c_1}}}.
\end{equation}
Under the same conditions the variational functional $Q$ takes the
form
\begin{equation}
\label{eq:8}Q = B(g){L_x}\frac{1}{{{L_y}}}{({c_2} - {c_1})^2},
\end{equation}
where now the constant $B$ depends on the geometry only. It should
be emphasized however, that the arguments supporting the two
equations above are mainly heuristic and based on the grounds of
irreversible thermodynamics and dimensional analysis. It is perhaps
interesting to notice that a formal proof of these equations is
lacking both in terms of random walks on a lattice or in continuous
time random walks.

In addition to the above toy-model fractal geometry, one might study
more complex objects, i.e. higher fractal generations. As an
illustration we present in Figure 2 the second and the fourth
fractal iterations of the fractal generator of Figure 1. These
irregular objects will play the role of counter-membranes in the
sequel.

\begin{figure}[h]
\vspace{-9pt}
        \centering
                \includegraphics[width=0.60\textwidth]{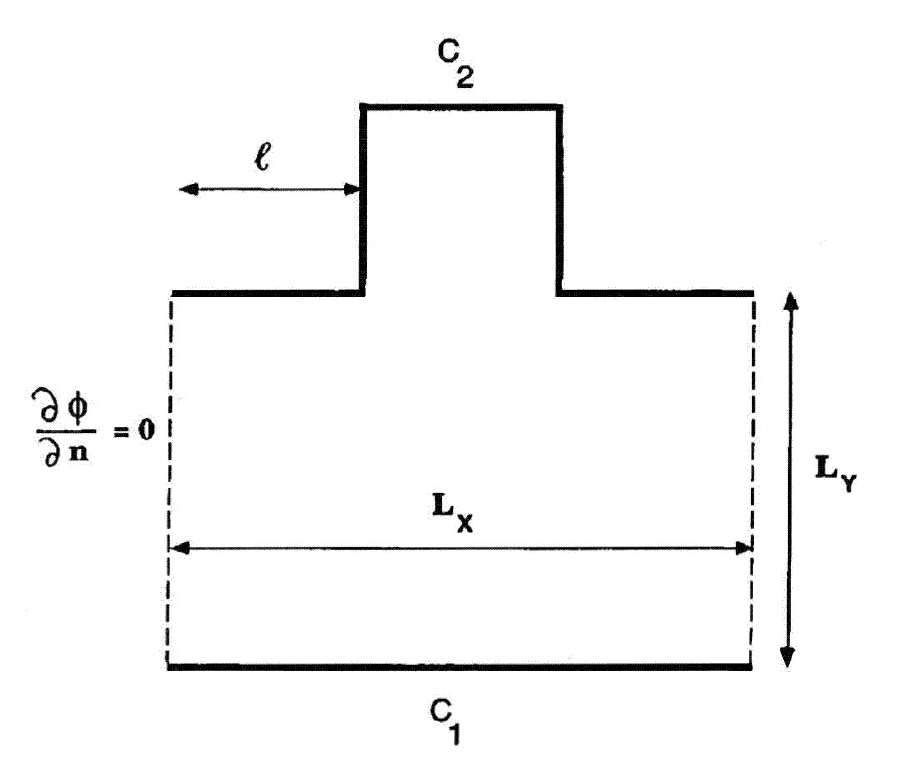}
                \vspace{-0.6cm}
                \caption{Schematic representation of the diffusion cell for the first fractal iteration  of a fractal generator.
                 The dimensions of the cell, ${L_x} = 3l$ and ${L_y} = 2l$, as well as the boundary conditions are indicated.
                 For this fractal ${D_F} = \frac{{\log 5}}{{\log 3}}.$}

\end{figure}
We now focus on the case of mild diffusion. The purpose is to
estimate the coefficients $A$ and $B$ analytically in this limit. To
this end we follow the so-called "independent field approximation"
(IFA). This is a coarse-graining argument based on
compartmentalization of the full continuous space in which diffusion
takes place into a finite number of properly selected
non-overlapping and non-interacting rectangular regions. Here, one
can write down closed expressions for the entropy production and the
variational functional assuming linear concentration profiles. The
nonlinear dependence of the field in terms of the spatial
coordinates arising from the irregularities of the boundaries is
thus approximated by piecewise linear functions, entailing
discontinuities of the equipotential lines at the boundaries between
the cells.

\begin{figure}[h]
\vspace{-15pt}
        \centering
                \includegraphics[width=1.20\textwidth]{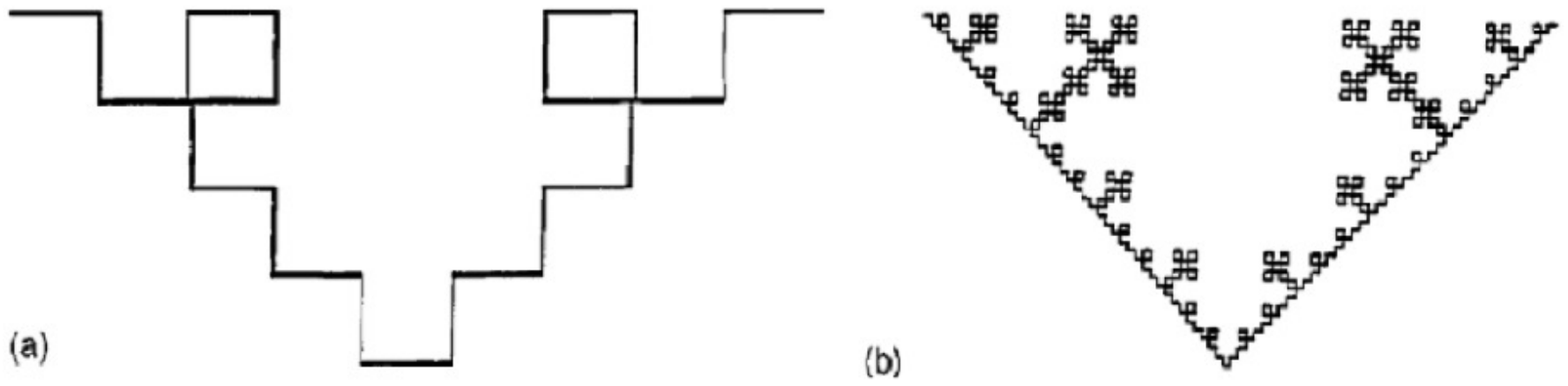}
                \vspace{-10.6cm}
                \caption{Shown are (a) the second and (b) the fourth fractal iterations of the  fractal generator of  Figure 1.}

\end{figure}

\begin{figure}[h]
\vspace{-10pt}
        \centering
                \includegraphics[width=1.20\textwidth]{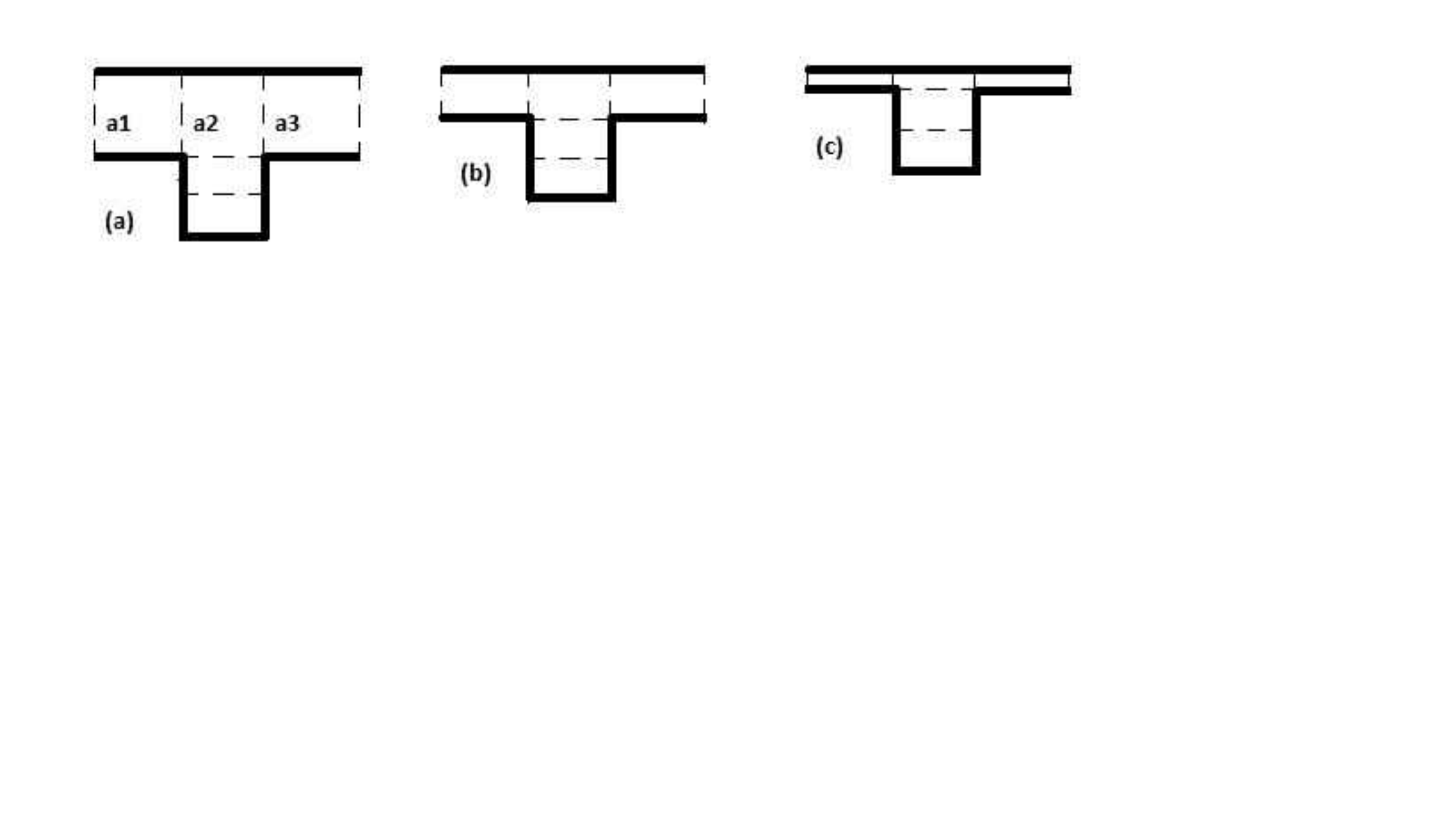}
                \vspace{-8.0cm}
                \caption{Schematic representation of the diffusion cell for the first fractal iteration of a fractal
                generator. The distance ${L_y}=ml$ (here $l=1$) is successively reduced from left to right: a) ${L_y}
                =m=0.3$ to b) ${L_y} =m= 0.1$ and c) ${L_y}=m \to 0$.}

\end{figure}
Let us illustrate how the IFA works for the first few fractal
generations of the cell. We separate the cell into the three
rectangles shown in Figure 3. Accepting  linear concentration
profiles in each of the side parts ($a_{1}$) and ($a_{3}$) and
applying eq.(5) with ${L_x} = l$, ${L_y} = ml$ (i.e. setting $F =
Dk({c_2} - {c_1})\ln \frac{{{c_2}}}{{{c_1}}}$) one finds for the
respective entropy production
\begin{equation}
\label{eq:9}{P_{a_{1}}} = {P_{a_{3}}} = \frac{F}{m}.
\end{equation}
Assuming furthermore a linear half penetration inside the pore (that
is, a linear concentration profile until the middle of the
pore$^{19-20}$, thereby neglecting the remaining active zone) the
central region ($a_{2}$) from eq.(5) with ${L_x} = l$ and ${L_y} =
ml + \frac{l}{2}$ reads
\begin{equation}
\label{eq:10}{P_{a_{2}}} = F\frac{l}{{ml + \frac{l}{2}}} =
F\frac{2}{{2m + 1}}.
\end{equation}
Thus, the total entropy production of the cell becomes
\begin{equation}
\label{eq:11}{P_{tot}} = {P_{a_{1}}} + {P_{a_{2}}} + {P_{a_{3}}} =
2F\frac{{3m + 1}}{{m(2m + 1)}}.
\end{equation}
Applying now eqs.(5) and (6) for the whole cell with ${L_x} = 3l$
and ${L_y} = ml$  we find (remember that we deal here with mild
diffusion, see eq.(7)) that
\begin{equation}
\label{eq:12}{P_{tot}} = \frac{{3FB(g)}}{m}.
\end{equation}
So that, for the first iteration (Figure 1) the coefficient $B$ is
of the form
\begin{equation}
\label{eq:13}{B^{(1)}} = \frac{{6m + 2}}{{3(2m + 1)}} = \frac{2}{3}
+ \frac{2}{3}\frac{m}{{2m + 1}}.
\end{equation}
Similarly, for the second fractal iteration (Figure 2a) we obtain
\begin{equation}
\label{eq:14}{B^{(2)}} = \frac{{324{m^2} + 150m + 12}}{{9(6m + 1)(6m
+ 3)}} = \frac{4}{9} + \frac{2}{3}m\left[ {\frac{1}{{2m + 1}} +
\frac{2}{{1 + 6m}}} \right],
\end{equation}
and for the third
\begin{equation}
\label{eq:15}{B^{(3)}} = \frac{8}{{27}} + \frac{2}{3}m\left[
{\frac{1}{{2m + 1}} + \frac{2}{{6m + 1}} + \frac{4}{{18m + 1}}}
\right],
\end{equation}
where the superscript $(i)$ denotes the $i-th$ fractal iteration.
Proceeding in the same lines one can find and prove by induction
that for the $i-th$ fractal iteration it holds
\begin{equation}
\label{eq:16}{B_m^{(i)}} = {\left( {\frac{2}{3}} \right)^i} + \left(
{\frac{2}{3}} \right)m\sum\limits_{k = 1}^i {\frac{{{2^{k - 1}}}}{{1
+ 2 \cdot {3^{k - 1}}m}}}.
\end{equation}
The superscript ($i$) denotes the corresponding fractal generation
(i.e.  $i=1$  is for the first fractal generation, $i=2$  is for the
second etc) and $m$ indicates the respective inter-electrode
distance.

Pausing for a moment the analysis of $B_m^{(i)}$, we investigate
some features for the total entropy production of the fractal
generations. Especially, one can easily observe that for the first
few fractal iterations the total entropy production is
\begin{equation}
\label{eq:17}P_{tot}^{(1)} = 2P_{a_1}^{(1)} +
P_{a_2}^{(1)},~~P_{tot}^{(2)} = 4P_{a_{1}}^{(2)}+2P_{a_2}^{(2)} +
P_{a_3}^{(2)},
\end{equation}
and
\begin{equation}
\label{eq:18}P_{tot}^{(3)} = 8P_{a_1}^{(3)} + 4P_{a_2}^{(3)} +
2P_{a_3}^{(3)} + P_{a_4}^{(3)},~~P_{tot}^{(4)} = 16P_{a_1}^{(4)} +
8P_{a_2}^{(4)} + 4P_{a_3}^{(4)} + 2P_{a_4}^{(4)}+P_{a_5}^{(4)},
\end{equation}
where the superscript ($i$) refers to the fractal iteration and the
index $a_{i}$ corresponds to the geometrical structure that one can
encounter in each fractal (see also Figure 3). For example, the
index $a_{1}$ (or equivalently $a_{3}$) refers to the left (right)
side part of the first fractal generation, while $a_{2}$ denotes the
central region. Similarly, the indices $a_{4}$, $a_{5}$ refer to the
central regions of the second and third fractal generations
respectively etc. The above procedure implies that for the $i-th$
fractal iteration the total entropy production should be written as
\begin{equation}
\label{eq:19}P_{tot}^{(i)} = {2^i}P_{a_1}^{(i)} + {2^{i -
1}}P_{a_2}^{(i)} + {2^{i - 2}}P_{a_3}^{(i)} + {2^{i -
3}}P_{a_4}^{(i)} + ...,
\end{equation}
or in a more compact form
\begin{equation}
\label{eq:20}P_{tot}^{(i)} = \sum\limits_{k = 0/i - k \geqslant
0}^\infty  {{2^{i - k}}} P_{k}^{(i)}.
\end{equation}
Here, the superscript ($i$) denotes the corresponding fractal
generation (i.e. $i=1$ for the first fractal generation, $i=2$ for
the second etc), whereas $k$ corresponds to each geometrical
structure that appears in the corresponding fractal generation as
explained above. Moreover, we can find that the entropy production
for each geometrical structure can be written as
\begin{equation}
\label{eq:21}P_{a_1}^{(i)} = \frac{F}{{{3^{i -
1}}m}}~~and~~P_k^{(i)} = \frac{{2F}}{{2 \cdot {3^{i - k}}m + 1}},k
\geqslant 1,i \geqslant k,
\end{equation}
while the above polynomials are connected through the recurrence
relations
\begin{equation}
\label{eq:22}\begin{array}{*{20}{c}}
  {\frac{{P_{k+1}^{(l)}}}{{P_{k}^{(l)}}} = 1 + \frac{{4 \cdot {3^l}m}}{{{3^{k + 1}} + 2 \cdot {3^l}m}},}&{l \geqslant k,}&{k \geqslant 1}
\end{array}~~and~~ \begin{array}{*{20}{c}}
  {\frac{{P_k^{(l)}}}{{P_{a}^{(l)}}} = \frac{{2 \cdot {3^{ - 1 + l + k}}}}{{{3^k} + 2 \cdot {3^l}m}}}&{l \geqslant k,}&{k \geqslant 1}
\end{array}.
\end{equation}
Note also that the total entropy production for each fractal
generation and geometrical structure can be written as
\begin{equation}
\label{eq:23}P_{tot}^{(i)} = F\frac{{{2^i}}}{{{3^{i - 1}}m}} +
2F\sum\limits_{k = 1}^i {\frac{{{2^{k - 1}}}}{{1 + 2 \cdot {3^{k -
1}}m}}}.
\end{equation}
Coming back to the investigation of the coefficients $B_m^{(i)}$  we
observe that the first derivative with respect to $m$ is positive
\begin{equation}
\label{eq:24}\frac{{\partial B_m^{(i)}}}{{\partial m}} =
\frac{2}{3}\sum\limits_{k = 1}^i {\frac{{{2^{k - 1}}}}{{\left( {2
\cdot {3^{k - 1}} + \frac{1}{m}} \right){m^2}}}},
\end{equation}
implying that $B_m^{(i)}$  is an increasing function of $m$. We
should also observe here that for m=  1, 2, 3 one finds again the
results of$^{32}$, (geometries for the Far field)
\begin{equation}
\label{eq:25}B_1^{(1)} = \frac{8}{9} \simeq 0.8889, B_1^{(2)} =
\frac{6}{7} \simeq 0.8571,...,B_1^{(5)} =
\frac{{245,451,914}}{{289,739,835}} \simeq 0.8471,
\end{equation}
with a limiting value (in the context of the IFA approximation)
\begin{equation}
\label{eq:26}B_1^{(10)} =
\frac{{22,614,533,817,495,831,212,892,038}}{{26,698,620,787,529,312,076,958,125}}
\simeq 0.8470.
\end{equation}

\section{Predictions for the Near Field}
As also pointed out in the introduction, the "Near Field"
 is a situation not studied before in the literature. Supposing that in the phenomenon
there is no new physics, we can apply our real space renormalization
scheme (IFA) in order to obtain predictions and compare them with
the finite element predictions, considered here as reference values.
More specifically, we can easily confirm that for $m = 0.3$ the
predictions of our real space renormalization scheme are
\begin{equation}
\label{eq:27}B_{0.3}^{(1)} = \frac{{19}}{{24}} \simeq
0.7916,B_{0.3}^{(2)} = \frac{{359}}{{504}} \simeq 0.7123, ...,
B_{0.3}^{(5)} = \frac{{6186317}}{{9069732}} \simeq 0.6821,
\end{equation}
approaching asymptotically the value (in the framework of the IFA)
\begin{equation}
\label{eq:28}B_{0.3}^{(10)} =
\frac{{2,574,446,468,051,339,779,667,687}}{{3,776,514,354,889,019,285,033,772}}
\simeq 0.6817.
\end{equation}
Similarly, for m=0.1 the predictions of the renormalization are
\begin{equation}
\label{eq:29}B_{0.1}^{(1)} = \frac{{13}}{{18}} \simeq
0.7222,B_{0.1}^{(2)} = \frac{7}{{12}} \simeq
0.5833,...,B_{0.1}^{(5)} = \frac{{37388}}{{73143}} \simeq 0.5112.
\end{equation}
Giving a limiting value in the context of the IFA approximation
\begin{equation}
\label{eq:30}B_{0.1}^{(10)} =
\frac{{16,308,004,933,962,338,624}}{{31,975,093,599,832,520,109}}
\simeq 0.5100.
\end{equation}
Quite curiously, the renormalization procedure gives also a finite
limit when $m \to 0$, that is
\begin{equation}
\label{eq:31}B_{0.0}^{(1)} = \frac{2}{3} \approx 0.667,B_{0.0}^{(2)}
= \frac{4}{9} \approx 0.444,B_{0.0}^{(3)} = \frac{8}{{27}} \approx
0.296, B_{0.0}^{(4)} = \frac{{16}}{{81}} \approx 0.198.
\end{equation}
Thus, we can conclude that $B_{m = 0}^{(i)} = {\left( {\frac{2}{3}}
\right)^i}$, as it is expected from the Cantor structure of the
counter membrane. Table I presents the values of $B_{m}^{(i)}$
predicted by the IFA for the first eight fractal iterations. It
appears that $B_{m}^{(i)}$ take intermediate values of the order of
0.5 to 0.9. A monotonic decay is observed for successive prefractal
generations as the distance $L_{y}=ml+l/2$ remains fixed. It is
clear that when $L_{y}$ becomes very large ($L_{y}\to \infty$)
$B_{m}$ tends to unity. As an empirical value coming now from our
IFA approximation when $L_{y}= 0.1$ the coefficient $B$ for the
limiting fractal object approaches the value 0.51 (see Tables I,II).
On the other hand, for a given fractal iteration as $L_{y}$
increases the corresponding values of $B$'s are increasing.
Especially, it can be easily shown (through the IFA) that $\mathop
{\lim }\limits_{m \to \infty } B_m^{(i)} = {\left( {\frac{2}{3}}
\right)^i} + \frac{1}{3}{\sum\limits_{k = 1}^i {\left( {\frac{2}{3}}
\right)} ^{k - 1}}$. From here, one can confirm that for large
distances (i.e. $m \to \infty $ or equivalently $L_{y}\to \infty$)
the limiting value of $B_m^{(i)}$ of every fractal iteration is
equal to unity, as the spatial distribution of the concentration
becomes roughly uniform for large distances ${L_y}$. Indeed, Table
II shows the limiting value of the coefficients $B_{m}$ as the
distance $L_{y}$ is increased. More specifically, we observe a
monotonic increase for the limiting value from 0.5 to unity as
$L_{y}$ becomes very large (of the order of ten, $L_{y}\sim10$). As
an illustration of the above discussion, Figure 4 depicts the
evolution of the coefficients $B_{m}^{(i)}(g)$ (predicted through
the IFA) with respect to different fractal iterations for various
distances $m$ between the electrodes. In each case we observe a
decay for increasing fractal iterations and a saturation of the
corresponding coefficient (after the sixth fractal generation).

\begin{table}[ht]
\caption{In the table below, we present the values corresponding to
the first few fractal generations of the coefficient
$B_{m}^{(i)}(g)$ for different distances from the membrane. We
clearly observe a saturation after the seventh fractal iteration. It
is concluded that the limiting value of $B_m^{(f)}$ for the final
fractal object is a function of the fractal dimension ${D_f}$  of
the object.} \centering

\begin{tabular}{c c c c c c c c c}    
\hline\hline Distance (m) & $B_{{m}}^{(1)}$& $B_{{m}}^{(2)}$ &
$B_{{m}}^{(3)}$ & $B_{{m}}^{(4)}$ & $B_{{m}}^{(5)}$ &
$B_{{m}}^{(6)}$ & $B_{{m}}^{(7)}$ & $B_{{m}}^{(8)}$
\\ [0.5ex]
\hline 0.1 & 0.7222 & 0.5833 & 0.5304 & 0.5150 & 0.5112 & 0.5103 & 0.5101 & 0.5100\\
0.3 & 0.7917 & 0.7123 & 0.6892 & 0.6834 & 0.6821 & 0.6818 & 0.6817 & 0.6817 \\
1.0 & 0.8889 & 0.8571  & 0.8493 & 0.8475 & 0.8471 & 0.8471  & 0.8470 & 0.8470 \\
2.0 & 0.9333 & 0.9162 & 0.9122 & 0.9113 & 0.9111 & 0.9111 & 0.9111 & 0.9111\\
3.0& 0.9524 & 0.9407 & 0.9380 & 0.9374 & 0.9373 & 0.9372 & 0.9372  & 0.9372 \\
[1ex] \hline\hline
\end{tabular}
\label{table:nonlin}
\end{table}

\begin{table}[ht]
\caption{The limiting value of the coefficient $B_{m}^{(i)}(g)$
predicted by the IFA is listed for various inter-electrode distances
($m$).} \centering

\begin{tabular}{c c}    
\hline\hline Distance ($m$)& Limiting value \\ [0.5ex]
\hline 0.1 & 0.5100 \\
0.2 & 0.6178 \\
0.3 & 0.6817 \\
0.5 & 0.7584 \\
0.7 & 0.8042 \\
1.0 & 0.8470 \\
2.0 & 0.9111\\
3.0 & 0.9372\\
10.0 & 0.9794\\
[1ex] \hline \hline
\end{tabular}
\label{table:nonlin}
\end{table}

\begin{figure}[h]
\vspace{-20pt}
        \centering
                \includegraphics[width=0.70\textwidth]{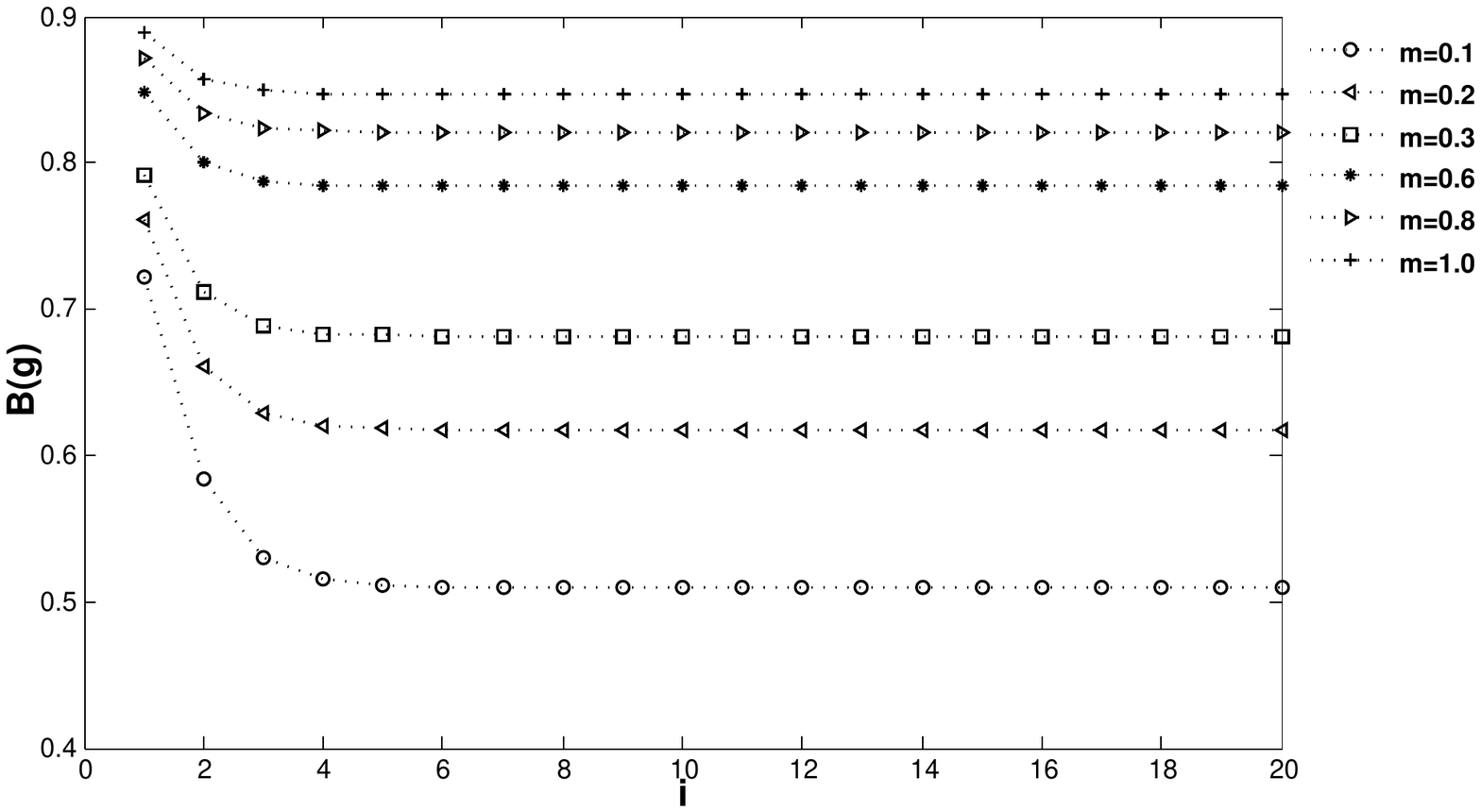}
                \vspace{-4.5cm}
                \caption{Shown are the coefficients $B(g)$ (eq.(16)) versus the fractal iteration $i$ for various
                inter-electrode ($m$) distances. We observe
                a saturation after the sixth fractal generation for a range of different distances.}
\end{figure}

\section{Irregular Boundaries: Numerical Approach in the Near Field}

The Laplace equation for geometries associated with the four first
generations has been solved numerically using an in-house developed
finite element code. All four geometries were discretized using
six-noded triangular elements with quadratic shape functions with
$C^{0}$ continuity at interelement boundaries. This resulted in a
quadratic interpolation of the potential field and a linear
variation of its gradient inside each element$^{34}$. The global
entropy production was then computed for various boundary conditions
by integrating the local entropy production obtained from the field
and the associated flux (essentially its gradient) by means of
relationship (2). This quadrature was implemented using a
25-integration points scheme derived by Laursen and Gellert$^{35}$.
Dirichlet boundary conditions with constant concentrations $c_{1}$
and $c_{2}$ are imposed to the models. The meshes used for the
calculations were produced using the GMSH meshing tool$^{36}$. A
very fine discretization of the studied domain was intentionally
selected in order to obtain sufficient precision on the values of
the local entropy production and of the functional for the purpose
of validation of the approximate coefficients. An element size on
the order of 4.5 $10^{-3}$ for the points of the model was selected
(for $L_{x}=1$) close to the fractal surface. The number of elements
and nodes considered for each generation is reported in Table III
(the min and max are related to the minimal and maximal distances
between the electrodes).

Using the above method one may estimate the factor $B(g)$ with a
very good precision (up to three significant digits). The numerical
coefficients (obtained from the aforementioned numerical method) are
presented in Tables IV-V while Table VI shows the comparison of the
theoretical (IFA) and the corresponding numerical values for the
coefficients $B_{{m}}^{(i)}$. These calculations indicate that the
proposed approximate method is satisfactory for the first few
fractal iterations and gets worse for increasing generation.
Especially, Tables IV-V illustrate the validation of the
phenomenological relation with the numerical evaluation via Finite
Elements, arguably the central result of this paper. From a careful
inspection of these Tables it is clear that as $L_{y}$ increases the
coefficients $B_{m}^{(i)}$ increases as well which is at least in
qualitative agreement with eq.(16) obtained by the IFA. We also
observe a decrease of the coefficients $A$'s and $B$'s as the
fractal generation increases. By inspecting Tables IV-V we observe
that for $L_{y}\to0.1$ a limiting value 0.65 is approached from the
Finite Elements method while the value 0.51 is predicted by the IFA
(see Table II). The latter indicates the failure of the IFA to be
accurate at high fractal generations. However, a qualitative
agreement between the IFA and the Finite Elements calculations is
evident, which means that for the final fractal object the
coefficients seem to tend to a limiting value independently of the
concentration $c_{2}$. Commenting now on Table VI which presents the
difference between the theoretical and numerical values (obtained
with Finite Elements) of the coefficients $B_{m}^{(i)}$ for
$c_{2}=0.9$, it is observed that as $L_{y}$ tends to zero
substantial deviations from the corresponding theoretical value
appear as a result of the divergence of the Laplacian field. Also,
for increasing fractal iterations, we see from the Table VI that the
relative error becomes important (see also below).

\begin{table}[ht]
\caption{The number of elements and nodes used for the numerical calculations (Finite Elements) for the first
four fractal generations. The corresponding values are related to the minimal and maximal distances between
the two electrodes.} \centering

\begin{tabular}{c c c}    
\hline\hline  Generation& Number of elements (min-max) &Number of nodes (min-max)
\\ [0.5ex]
\hline $ 1 $ & 118953-318347 & 239450-638544 \\
$ 2 $ &174147-385141  &350578-772872 \\
$ 3 $ & 199691-408749 & 402816-821238 \\
$ 4 $ &219401-428755  & 444236-863250 \\
[1ex] \hline\hline
\end{tabular}
\label{table:nonlin}
\end{table}

\begin{table}[ht]
\caption{The coefficients $B_{{m}}^{(i)ph}$ and $A_{{m}}^{(i)ph}$
(for the first four fractal generations) obtained from the
respective numerical values of the variational functional
$Q_{{m}}^{(i)ph}$ and the phenomenological entropy production
$P_{{m}}^{(i)ph}$, with concentration ${c_2} = 0.9$.} \centering

\begin{tabular}{c|| c c c c|| c c c c }    
\hline\hline  ${L_y}$& ${B^{(1)ph}}$& ${B^{(2)ph}}$ & ${B^{(3)ph}}$
& ${B^{(4)ph}}$ & ${A^{(1)ph}}$ & ${A^{(2)ph}}$ & ${A^{(3)ph}}$ &
${A^{(4)ph}}$
\\ [0.5ex]
\hline $ 0.1 $ &0.7780  & 0.6883 & 0.6620 & 0.6543 & 0.7779 & 0.6881 &0.6622  &0.6546 \\
$ 0.3 $ & 08664 & 0.8276  & 0.8170 & 0.8134 & 0.8665 & 0.8276  &0.8172 & 0.8134 \\
$ 0.5 $ &0.9066  &0.8833  &0.8766  & 0.8733 & 0.9064 & 0.8827 &0.8763  &0.8732  \\
$ 1.0 $ & 0.9500 & 0.9366 & 0.9333 &0.9300  & 0.9491 & 0.9364 & 0.9333 &0.9301 \\
[1ex] \hline\hline
\end{tabular}
\label{table:nonlin}
\end{table}

\begin{table}[ht]
\caption{The coefficients $B_{{m_i}}^{(i)ph}$  and
$A_{{m_i}}^{(i)ph}$ (for the first four fractal generations)
obtained from the respective numerical values of the variational
functional $Q_{{m_i}}^{(i)ph}$ and the entropy production
$P_{{m_i}}^{(i)ph}$, with concentration  ${c_2} = 0.01$.} \centering

\begin{tabular}{c|| c c c c|| c c c c }    
\hline\hline  ${L_y}$& ${B^{(1)ph}}$& ${B^{(2)ph}}$ & ${B^{(3)ph}}$
& ${B^{(4)ph}}$ & ${A^{(1)ph}}$ & ${A^{(2)ph}}$ & ${A^{(3)ph}}$ &
${A^{(4)ph}}$
\\ [0.5ex]
\hline $ 0.1 $ & 0.7779 & 0.6882 & 0.6620 & 0.6544 & 0.7776 & 0.6874 &0.6606  &0.6520 \\
$ 0.3 $ &0.8662 & 0.8280  & 0.8174 & 0.8139 &0.8661  & 0.8277  &0.8166  & 0.8128 \\
$ 0.5 $ &0.9070  & 0.8829 & 0.8758 & 0.8734 & 0.9069 &0.8827  & 0.8755 & 0.8728 \\
$ 1.0 $ & 0.9491 & 0.9361 & 0.9322 & 0.9308 & 0.9490 & 0.9360 & 0.9320 &0.9306 \\
[1ex] \hline\hline
\end{tabular}
\label{table:nonlin}
\end{table}

\begin{table}[ht]
\caption{Percentage difference, with concentration ${c_2} = 0.9$ and
various distances $L_{y}$, between the two electrodes for the
theoretical and the corresponding numerical values of the
coefficients $B_{{m}}^{(i)}$.} \centering
\begin{tabular}{c c c c c }    
\hline\hline  ${L_y}$& ${B^{(1)}}$& ${B^{(2)}}$ & ${B^{(3)}}$ &
${B^{(4)}}$
\\ [0.5ex]
\hline $ 0.1 $ &$7.2\%$  &$15.2\%$  &$19.9\%$  &$21.3\%$ \\
$ 0.3 $ &$8.6\%$  &$13.9\%$   &$15.7\%$  &$15.9\%$ \\
$ 0.5 $ & $ 8.1\%$ & $11.9\% $  & $12.9\%$ &$13.0\%$ \\
$ 1.0 $ &$6.4\%$ &$8.5\%$  &$9.0\%$  &$8.9\%$ \\
[1ex] \hline\hline
\end{tabular}
\label{table:nonlin}
\end{table}

Another attempt to improve the above procedure is to insert the
phenomenological (numerical) values for the coefficients
$B_m^{(1)ph}$, and actually obtain a "renormalized" value for the
higher order coefficients e.g. $B_m^{(2)}$ etc, thus improving
considerably the predictions of the theory. Combining the empirical
result for the first generation with a IFA half penetration argument
inside the new pores for the second fractal generation, one can find
\begin{equation}
\label{eq:32}B_{0.5}^{(2)ren} = \left( {0.908 - \frac{2}{3}}
\right)3l + 2\frac{8}{9}l = 0.834,
\end{equation}
while
\begin{equation}
\label{eq:33}B_{0.3}^{(2)ren} = 0.786~~and~~B_{0.1}^{(2)ren} =
0.639.
\end{equation}

Comparing now the above results we can find that $B_{0.5}^{(2)ren}$
differs from $B_{0.5}^{(2)ph}$ by only $5.8\% $. Similarly
$B_{0.3}^{(2)ren}$ differs from $B_{0.3}^{(2)ph}$ by only $5.7\% $
and $B_{0.1}^{(2)ren}$ differs from $B_{0.1}^{(2)ph}$ by $8.6\% $.
In the same way, we can proceed with the remaining orders. Also, one
can try various renormalization schemes together with the
theoretical arguments. The best results are obtained when one uses
for the prediction of the coefficient of the $k+1$ fractal iteration
${B^{(k + 1)}}$ only the empirical coefficients of the previous $k$
iteration and no theoretical argument at all. For instance, for the
case of the second fractal iteration for distance $m=0.5$ one can
find
\begin{equation}
\label{eq:34}B_{0.5}^{^{(2)fren}} = \left( {0.908 - \frac{2}{3}}
\right)3l + \frac{2}{3}0.949 \simeq 0.8746,
\end{equation}
where the superscript "fren" stands for "fully renormalized". Here,
the difference between $B_{0.5}^{^{(2)fren}}$ and
$B_{0.3}^{(2)ph}$ is of the order of $1.2\% $.
 Figures 5-12 present the spatial dependence of the entropy production for the first four fractal
iterations, for ${L_y} = 0.3$, for the cases of strong (${c_2} =
0.1$) and mild (${c_2} = 0.9$) diffusion respectively. We observe
that the behaviour of the local entropy production depends strongly
from $c_{2}$. In particular, for the case of strong diffusion (see
Figures 5,7,9,11) the appearance of a passive zone is observed
irrespectively of the fractal generation. In contrast, for the case
of mild diffusion the above statement breaks down, being indicated
by the spreading of the local entropy production near the
geometrical irregularities (see Figures 6,8,10,12).
\begin{figure}[h]
\vspace{-15pt}
        \centering
                \includegraphics[width=0.90\textwidth]{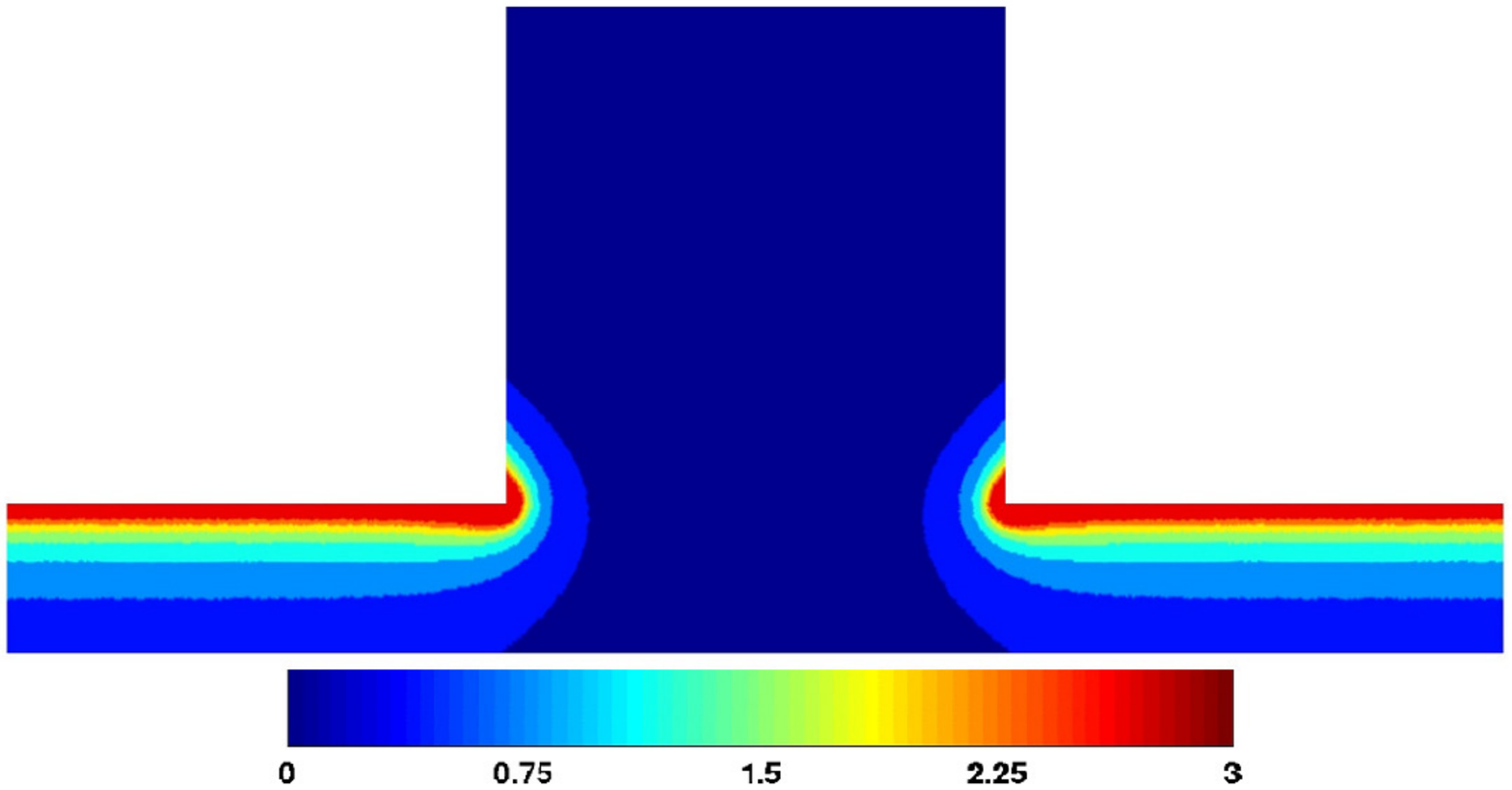}
                \vspace{-4.0cm}
                \caption{Spatial dependence of the entropy production for the diffusion cell of Figure 1. It is shown the case of
                   strong diffusion with ${c_2} = 0.1$, while the distance between the electrodes is ${L_y} = 0.3$. (Computer units.)}
\end{figure}

\begin{figure}[h]
\vspace{-15pt}
        \centering
                \includegraphics[width=0.90\textwidth]{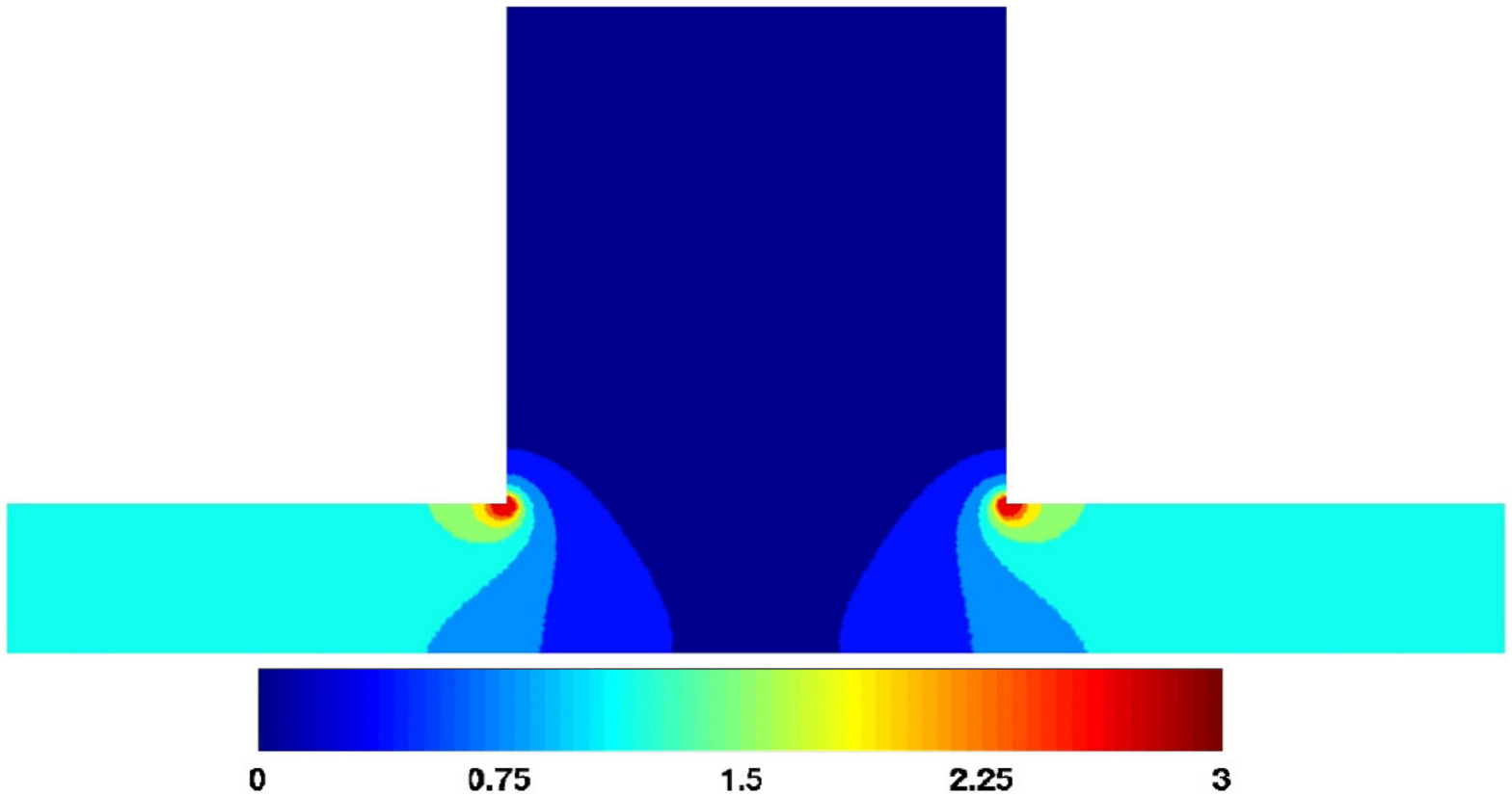}
                \vspace{-4.0cm}
                \caption{Spatial dependence of the entropy production for the diffusion cell of Figure 1. It is shown the case of
                   mild diffusion with ${c_2} = 0.9$, while the distance between the electrodes is ${L_y} = 0.3$. (Computer units.)}
\end{figure}

\begin{figure}[h]
\vspace{-15pt}
        \centering
                \includegraphics[width=0.90\textwidth]{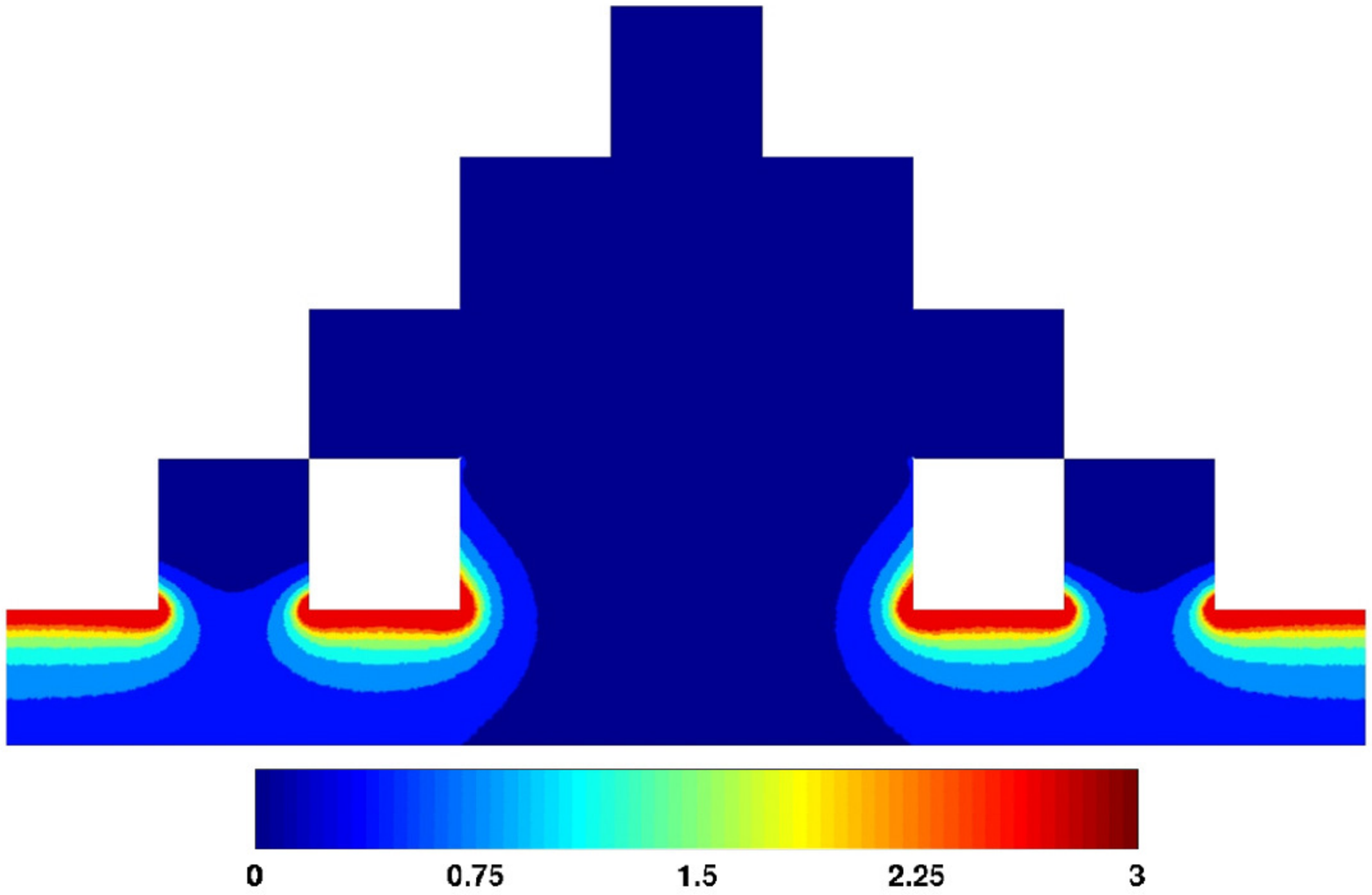}
                \vspace{-3.0cm}
                \caption{Spatial dependence of the entropy production for the second fractal iteration. It is shown the case of
                   strong diffusion with ${c_2} = 0.1$, while the distance between the electrodes is ${L_y} = 0.3$. (Computer units.)}
\end{figure}
Let us briefly recall here the argument which relates the
coefficients $A$ and $B$ through the active zone concept for the
case of mild diffusion, i.e. $\left| {({c_2} - {c_1})/{c_1}} \right|
= O(\varepsilon )$, $\varepsilon  \ll 1$, with ${c_1} = O(1)$. Here,
from eqs (7) and (8) one can conclude that $P \approx {\varepsilon
^2}$  and $Q \approx {\varepsilon ^2}$ while $B(g) \approx \mathop
{\lim }\limits_{{c_2} \to {c_1}} {A^{mild}}({c_1},{c_2};g)$ up to a
factor of $Dk/{c_1}$. The basic idea of this argument is that the
profiles of the concentrations are essentially linear in space in
the whole active volume and the nonlinearities are smoothly
approximated.

In the opposite situation, we have the strong diffusion. One
particular case of strong diffusion, which we study here in detail,
is the case of the harmonic measure (${c_1} = 1$, ${c_2} \to 0$).
\begin{figure}[h]
\vspace{-15pt}
        \centering
                \includegraphics[width=0.90\textwidth]{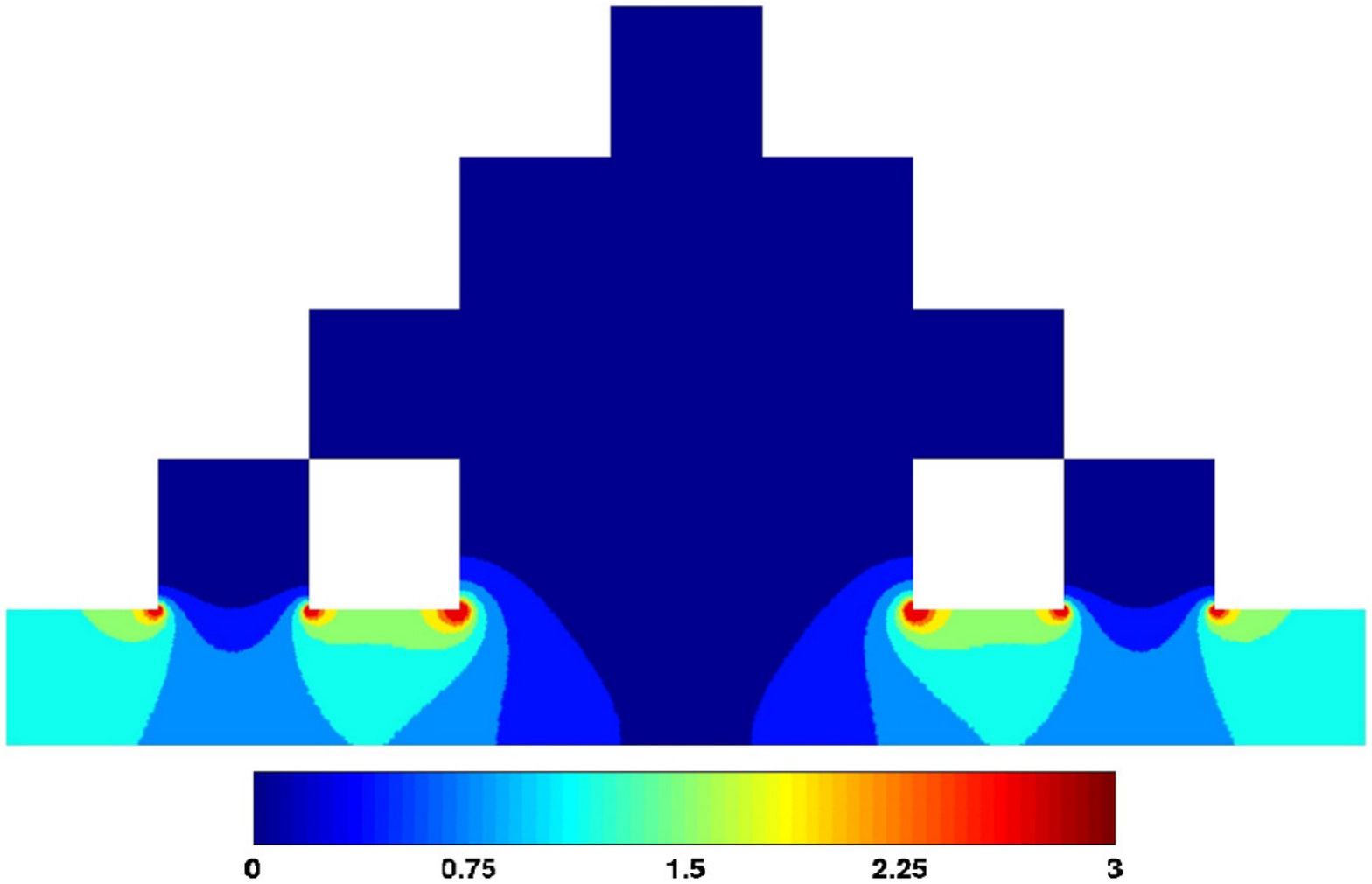}
                \vspace{-3.0cm}
                \caption{Spatial dependence of the entropy production for the second fractal iteration. It is shown the case of
                   mild diffusion with ${c_2} = 0.9$, while the distance between the electrodes is  ${L_y} = 0.3$. (Computer units.)}
\end{figure}
\begin{figure}[h]
\vspace{-15pt}
        \centering
                \includegraphics[width=0.90\textwidth]{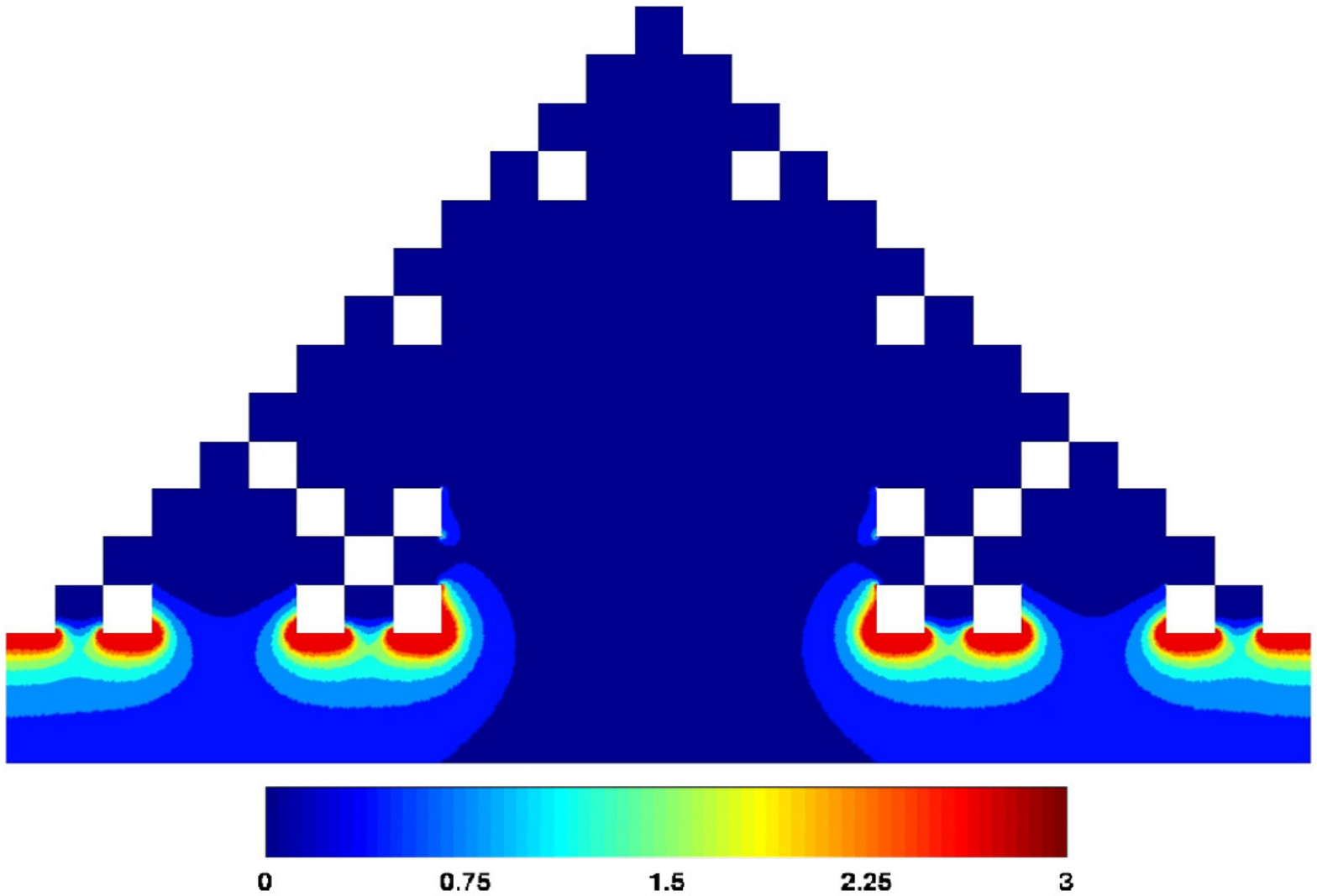}
                \vspace{-2.5cm}
                \caption{Spatial dependence of the entropy production for the third fractal generation. It is shown the case of
                   strong diffusion with ${c_2} = 0.1$, while the distance between the electrodes is  ${L_y} = 0.3$. (Computer units.)}
\end{figure}
\begin{figure}[h]
\vspace{-15pt}
        \centering
                \includegraphics[width=0.90\textwidth]{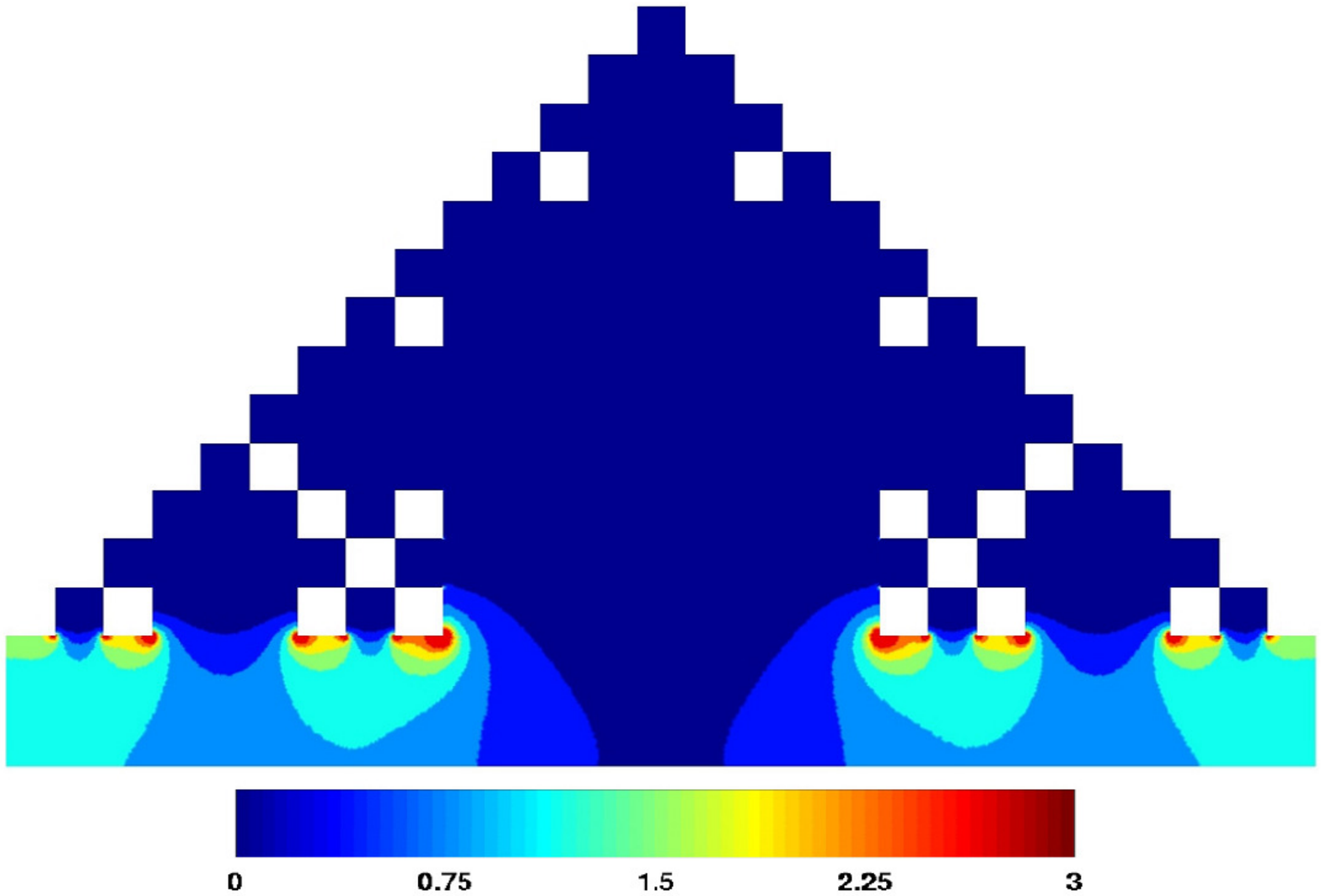}
                \vspace{-2.5cm}
                \caption{Spatial dependence of the entropy production for the third fractal generation. It is shown the case of
                   mild diffusion with ${c_2} = 0.9$, while the distance between the electrodes is ${L_y} = 0.3$. (Computer units.)}
\end{figure}
\begin{figure}[h]
\vspace{-15pt}
        \centering
                \includegraphics[width=0.90\textwidth]{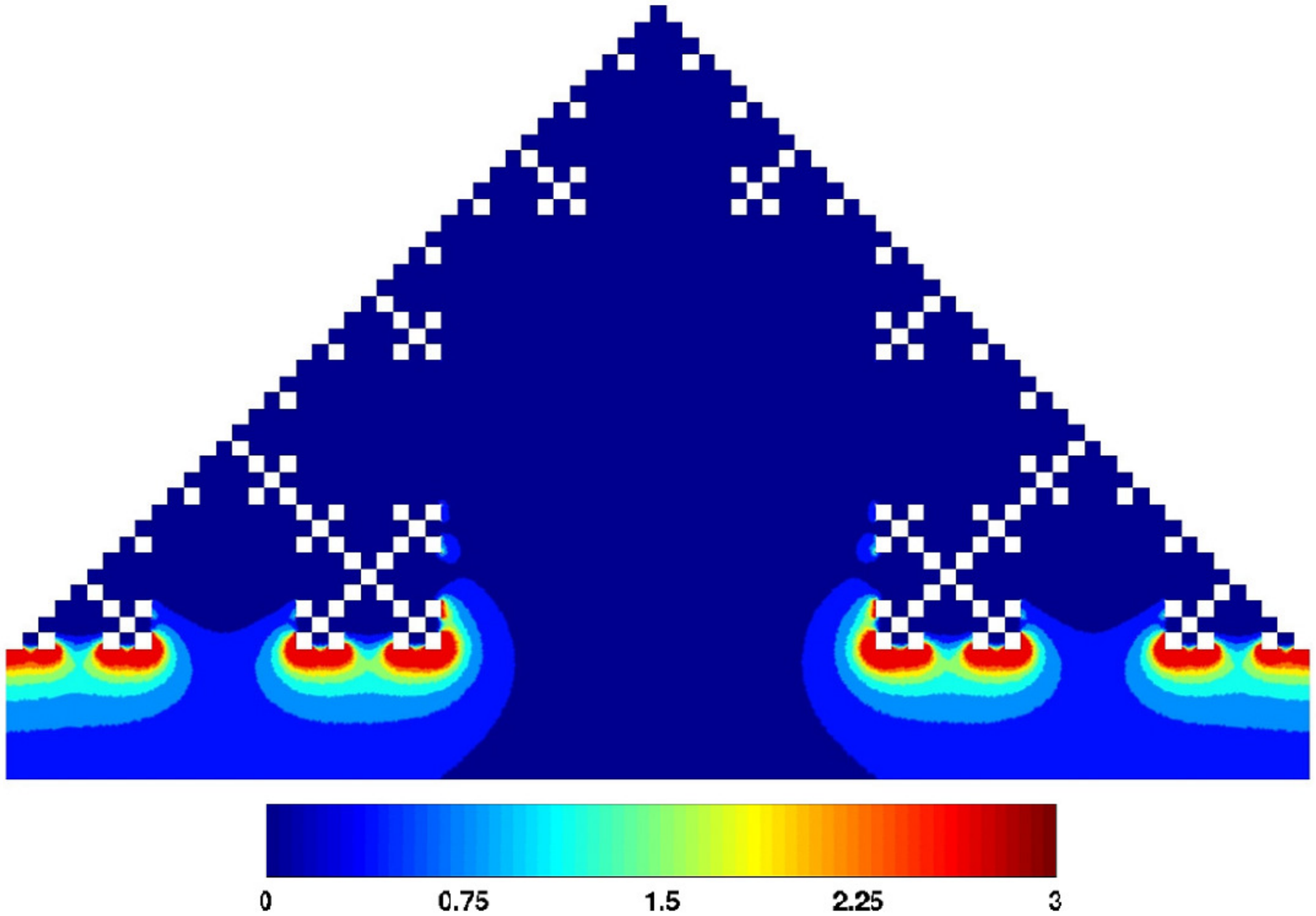}
                \vspace{-2.5cm}
                  \caption{Spatial dependence of the entropy production for the fourth fractal generation. It is shown the case of
                   strong diffusion with ${c_2} = 0.1$, while the distance between the electrodes is ${L_y} = 0.3$. (Computer units.)}
\end{figure}
\begin{figure}[h]
\vspace{-15pt}
        \centering
                \includegraphics[width=0.90\textwidth]{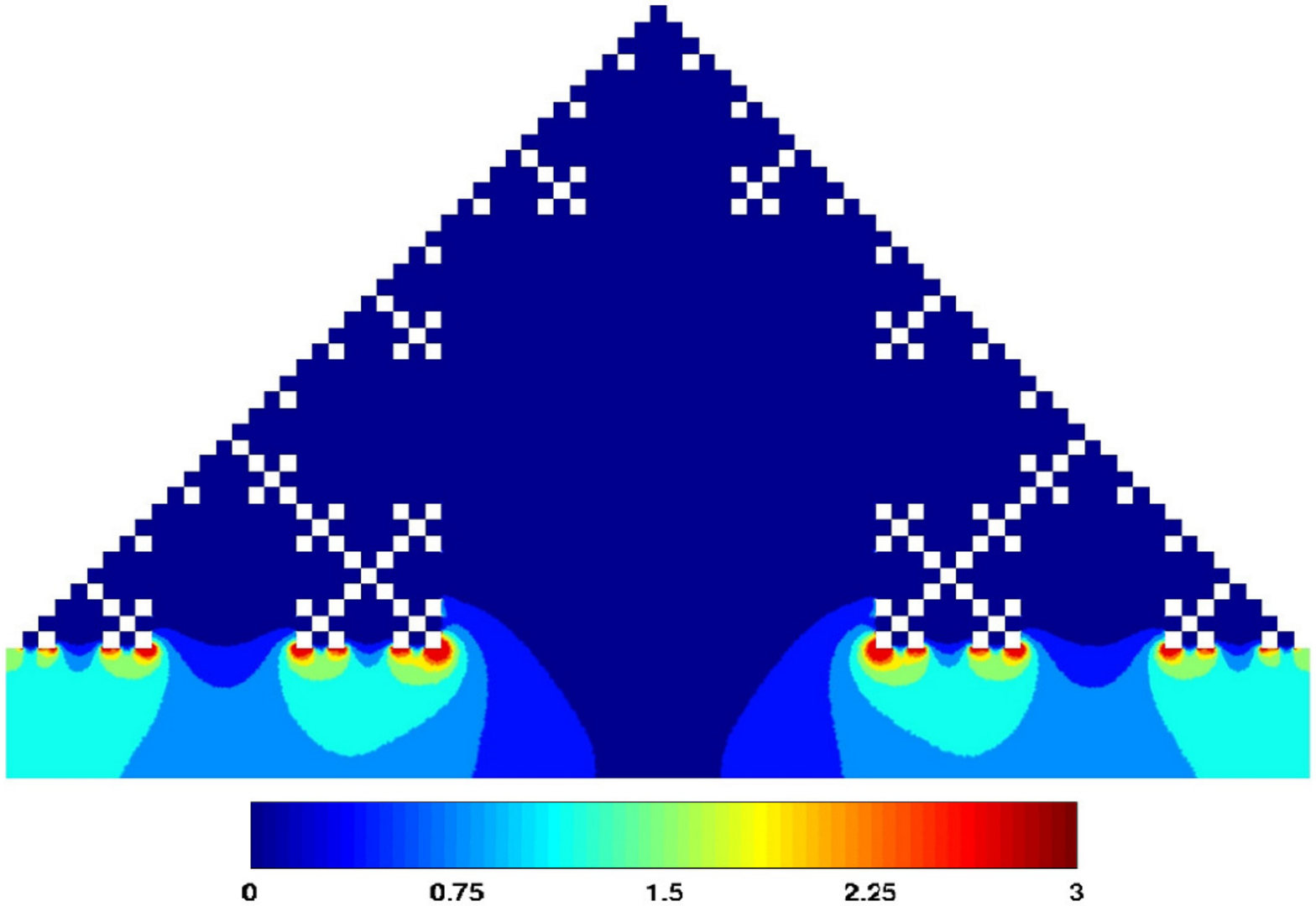}
                \vspace{-2.5cm}
                \caption{Spatial dependence of the entropy production for the fourth fractal generation. It is shown the case of
                mild diffusion with ${c_2} = 0.9$ and distance between the electrodes ${L_y} = 0.3$. (Computer units.)}
\end{figure}
Exactly in this case we obtain a visual interpretation of Makarov's
theorem. The measure is spreading on the surface of the working
membrane almost uniformly in a length which roughly coincides with
the diameter of the active membrane $l + \frac{l}{2} + \frac{l}{2} +
l = 3l$. In this case, Makarov's theorem guarantees that at least
for ${c_2} > 0.01$, $A=B$ holds (see also Tables IV-V). We thus
obtain a very important result of our real space renormalization
scheme: Estimating the values of $B(g)$ with the active zone
approximation (Mandelbrot's argument) we obtain also the value of
$A$ in the whole range of ${c_2}$, that is in all cases except from
the case where ${c_2} < 0.01$.

The above remarks indicate that the aforementioned IFA scheme is
valid (in the Far and Near domain) at least for the first few
fractal iterations for both strong and mild diffusion. Indeed, it is
remarkable that such a simple renormalization scheme gives accurate
predictions for the first fractal generations and leads at least to
qualitative results for higher iterations (here up to the fourth).
From our numerical calculations it turns out that the quality of the
IFA prediction gets worse as the fractal generation increases and/or
as the inter-electrode distance tends to zero (see Table VI). The
basic reason behind this result is that the IFA is a coarse-graining
approximation which neglects the fine microscopic details of the
fractal objects. This is exactly the way to the passive zone
approximation. Thus, it is a challenging open question whether the
above conclusions can be straightforwardly generalized for even
higher fractal iterations, where the IFA approximation is expected
to fail. This constitutes a heavy computational task especially for
high enough fractals iterations. In particular, three questions
remain to be checked: a) to what extent the fully renormalized
scheme is valid for higher fractal iterations, b) whether the
limiting value predicted by the IFA approximation holds as a
qualitative argument and c) the investigation of a better
renormalization scheme that takes into account the respective
correlations in the distribution of the harmonic measure especially
for $m\rightarrow0$.

\section{Conclusions and Discussion}

The transport associated to molecular transfer of neutral particles
from a diffusional field to an irregular membrane is examined from
the standpoint of irreversible thermodynamics. Combining
non-equilibrium thermodynamics and an argument developed initially
by Mandelbrot et al.$^{19}$, and then by Makarov et al.$^{28,29}$,
one is led to probe and predict the entropy production and the
variational functional of the transfer. Detailed predictions are
given for the whole range of distances from the Far field to the
Near field corroborated by finite differences$^{2,32,33}$ and Finite
Elements calculations.

The central idea of the active zone concept$^{2}$ is that on the
grounds of the Laplacian transfer on a membrane, certain parts of
the membrane remain passive and all thermodynamic properties of the
membrane can be estimated with a very good approximation by the
contribution of the "active parts". In the context of the active
zone introduced by spectroscopic impedance, a system with irregular
boundaries possesses,  in the Makarov regime where the reactive
system is "almost entirely diffusive",  the same impedance as a
smaller system with regular boundary conditions. This
${L_{app.Mak}}$ "apparent, spectroscopic length",  is not the same
as the quantity ${L_{app,entr}} = B(g){L_x}$  but the conclusion
remains the same: an entirely diffusive system with irregular
boundary conditions possesses the same entropy production as a
system with smaller diameter and regular (flat) boundary conditions.
This can be seen as an interpretation of Makarov's theorem in terms
of irreversible thermodynamics. Do these remarks apply to the "Near
field" ? In the Near field the original Mandelbrot's argument boils
down and the notion of spectroscopic impedance is relaxed. However,
as it is shown here one does not expect "new physics" and
irreversible thermodynamics is still valid. Our remarks, are
corroborated from the evaluation of the local entropy production
around the von Koch curve in the Near field.

To be concrete, the entropy production and the variational
functional of Laplacian transfer for geometries associated with the
first few fractal iterations leading eventually to a fractal
generator (von Koch curve) are studied. We first recover and
recapitulate results for the Far field. Then, we turn to the case of
the "Near field", a situation not studied before in the literature,
for both strong and mild diffusion. Detailed charts of the local
entropy production around the first four iterations are given, and
the global entropy production is investigated carefully. From these
studies it is clear that the Laplacian fields are in a sense a
"smooth measure" around irregular objects. Moreover, the analytic
procedure which allows one to approximate the entropy production
seems to be quite powerful and straightforward. Although the spatial
distribution of the local entropy production changes greatly from
the mild to strong diffusion, the IFA analysis and the global
entropy production obey the same rules in the Far and in the Near
field. A fair evaluation of the IFA arguments implies that the
difference between the predicted and the numerically obtained values
increases with the fractal iterations. An interpretation of the
above statement is that the IFA is a coarse-graining approximation
which neglects the fine microscopic details of the fractal objects
(passive zone approximation). More specifically, in the framework of
our IFA renormalized $B$'s for the first three generations differ by
$5-15\%$ from the numerical reference values, and the fully
renormalized B's coincide up to $2\%$ difference to the numerical
values. As already pointed out, the generalization of these results
for higher fractal iterations constitutes an interesting open
problem. Another question that arises from our investigation is
whether the predicted by the IFA limiting value of the coefficient
$B$ for high fractal generations holds at least qualitatively. This
task requires heavy computations for the Laplacian transfer of high
fractal objects and has to be performed in the context of a
supercomputer. Finally, as we have also mentioned in the past, our
predictions on such dissipation trends should be amenable to
experimental testing in a suitably constructed diffusion cell.

{}

\end{document}